\documentstyle[12pt]{article}
\title{
Hybrids  with heavy quarks:\\
 from potential to string
}
\vspace{2cm}
\author{
Yu.B.Yufryakov\thanks
{e-mail:yufryakov@vxitep.itep.ru}
\\Institute of Theoretical and Experimental Physics\\117259,
Moscow}
\date{}
\newcommand{\be}{\begin{equation}}
\newcommand{\ee}{\end {equation}}
\begin{document}
\maketitle

\begin{abstract}

 We consider hybrid states with heavy quarks in the frame of recently
proposed constituent gluon model. Limiting regimes for model
Hamiltonian are discussed. It is shown that these regimes match
smoothly, hence predictions of the model are definite enough. We
predict lowest $c \bar c g$ hybrids at $4.1\pm0.1$ GeV and $b \bar b g$
hybrids at $10.5\pm 0.1$ GeV. Regge-trajectories for hybrids are
derived from the formalism. Experimental aspects of our predictions are
discussed.
\end{abstract}

\newpage

 One of the main challenges to experimentalists in hadron physics is
that existence of gluonic degrees of freedom had not been confirmed yet
by direct experiment in the nonperturbative regime. Presence of gluonic
fields in the QCD Lagrangian make us think that they should reveal
themselves as a constituent particles in the form of glueballs and
hybrids. The important questions are : how can we distinguish an object
containing valent gluon from ordinary meson and baryon resonances and
where can we find this object ? Hybrids have more attractive features
(with respect to glueballs - see [1] ). Most of models predict hybrids
with light quarks at 1.5-2 GeV , but there are too many resonances at
this region and experimental situation is far from obvious (despite the
fact that standard nonets are overpopulated now). On the other hand ,
charmonium and bottomonium spectroscopy is well understood now and one
can identify heavy exotic much easier than the light one. These are the
reasons for intensive studies of heavy quark hybrids  carrying
now in the literature [2].

 Recently constituent gluon model based on Vacuum Background Correlators
method [3] and QCD string model [4,5] was proposed [6,7]. This model was
applied to heavy hybrids spectroscopy in [8] where masses of lowest and
states were estimated. In this letter we discuss potential and string
regimes for QCD string in the case of heavy hybrids.

 Consider adiabatic approximation for heavy hybrids. For beginning we
neglect the Coulomb interaction. Then it is natural that "fast" gluonic
subsystem energy is of order $\sqrt \sigma$ ($\sigma$ is string tension)
and quark subsystem energy should be $\sqrt \sigma$ times square root
from the mass ratio i.e. $E_{q\bar q}\sim \sqrt \sigma \sqrt{\sqrt
\sigma/m}$ where $m$ is heavy quark mass. For short distances interquark
potential must be oscillator one. It's easy to see that unharmonic
corrections will be of the next order on adiabatic expansion parameter
 [8]. ( For interquark distance $\rho$ much larger than $1/\sqrt \sigma$
interquark potential should be $\sigma \rho$, of course.) Therefore
gluon subsystem energy can be obtained after substraction of the
 oscillator interquark potential from the full Hamiltonian.

 Coulomb interaction almost doesn't change nothing. Interquark Coulomb
repulsion is negligible because its parameter $\alpha_s /6$ is very
small. Coulomb quark-gluon attraction change gluon energy more
drastically but still gluon subsystem energy is proportional to
 $\sqrt \sigma$ .

 Hamiltonian of the gluonic subsystem depending on $\rho$ as a parameter
(neglecting spin degrees of freedom) has been obtained in [8]. In short,
main assumptions made  are :1) Quenched approximation for quarks and
gluon ; 2) Area law for Wilson loops ; 3) Forbidden time backtracking
for all particles [4]; 4) Straight-line ansatz for minimal surface in
the area law [4]; 5) S-wave  for  quarks [8].The resulting Hamiltonian
reads (in Minkowski space-time ) :
$$
H_{gl}=\frac{p_r^2}{2 \mu_g}+\frac{\mu_g+a_1+a_2}{2}+\frac{{\vec L}
_{gl}^2}{2(\mu_g+\Pi_1+\Pi_2)}+\frac{\sigma^2}{2}(d_1 r_1^2+d_2 r_2^2)+
$$
\be
+\frac{\Pi {\vec L}_{gl}^2 \rho^2}{12{(\mu_g+2 \Pi)}^2 r^4}
-\frac{\Pi p_r^2 \rho^2}{6 \mu_g^2 r^2}
\ee
where
$$ a=\int_0^1 d\beta \nu (\beta) ; d=\int_0^1\frac{d\beta}{\nu(\beta)}
; \Pi=\int_0^1 d\beta \nu(\beta){(\beta-1)}^2
$$

 Here $\rho$ is interquark radius ; $\vec r$ is the  gluon 3-coordinate
 with respect to quarks c.o.m. (quarks have the same mass) ; $\vec r_1$
 and $\vec r_2$ are gluon-quark and gluon-antiquark 3-coordinates ;
$\beta$ parametrize the space part of the minimal surface (see [4]) ;
$\nu_1,\nu_2$ and $\mu_g$ are auxiliary fields ( effective string
energies and gluon mass) ; $L_{gl}$ is the angular momentum of gluon -
it is integral of motion in the  limit $\rho \to 0$. Strictly say,
$L_{gl}$
is not integral of motion cause condition $\rho <<r$ is not satisfied
for c- and b-quarks ( they are not enough heavy - for our purposes ).
We wish to emphasize that the gross structure of hybrid spectrum is
obtained in the limit $\rho \to 0$ ; $\rho^2/r^2$ terms are corrections
of order 400 MeV for charmonium hybrids and about 200 MeV for
bottomonium ones. For real hybrids angular momentum is distributed
anyway between quark and gluon subsystem , but general structure of
spectrum remains unchanged.

 In the limit $\rho \to 0$ one can let $\nu_1=\nu_2$ and obtain gluon
subsystem Hamiltonian in zeroth approximation :
\be
H_{gl}=\frac{p_r^2}{2 \mu_g}+\frac{\mu_g}{2}+\frac{L_{gl}^2}{2{(\mu_g
+2\Pi)}^2 r^4}+d\sigma^2 r^2+a
\ee

 This is a Hamiltonian of heavy-light string system described in [5]
with effective string tension $2\sigma$ - two quark-gluon strings merge
into the only string. There are two opposite limits for Hamiltonian (2)
described in [4,5]. Namely, for $L_{gl}=0$ one can integrate over
auxiliary fields $\mu_g$ and $\nu$ in the path integral representation
for $q \bar q g$ Green function (see [4,5]) and obtain :
\be
{\hat H}_{gl}=\sqrt{p_r^2}+2\sigma r
\ee
This is a "naive" Hamiltonian for zero-mass particle confined by scalar
potential $\sigma r$ (potential limit appeares also for constituents
much heavier than $\sqrt \sigma$ ). The opposite limit is string limit
when $r$ is not dynamical variable - this is a limit $L>>1$. Then one
is to replace $r$ by its extremal value :
\be
r^2=\frac{1}{2\sigma}\sqrt{\frac{L_{gl}^2}{\Pi d}}
\ee
and neglect $\mu_g$ as it is usual [4]. After that one can integrate
over field $\nu$ replacing  it  by  its  extremal  value :
\be
\nu=\left(\frac{2\sigma \sqrt{L(L+1)}}{\pi}\right )\frac{1}{\sqrt{1-
{(\beta-1)}^2}}
\ee
with the result :
\be
E=\sqrt{2\pi \sigma \sqrt{L(L+1)}}
\ee

  So, as it was shown in [5], string regime for Hamiltonian (2) leads
to Regge trajectories with the correct slope. For our case with
effective string tension $2\sigma$ the slope of Regge trajectories is
$2\pi\sigma$. Note that result (6) is obtained by means of adiabatic
approximation, hence it is valid for
\be
1<<L_{gl}<<m/\sqrt\sigma
\ee
Condition (7) is satisfied only for $b\bar b g$ hybrids and results
(4)-(6) are applicable only in this case. We do not consider pure string
case with $L_{gl}>>1$ when even heavy quark mass is negligible with
respect  to  string mass (6).

 These marginal regimes are purely academical . Indeed, gluon can't
have $L_{gl}=0$ cause physical states for gluon are electric and
magnetic (see [7]). On the other hand, here we are interested only with
lowest  states with $<L_{gl}^2>=2$ and do not evaluate masses of highly
excited states.

 In [2] variational procedure for the auxiliary field $\nu$ was used.
In this letter we'll show  that {\bf both} potential and string regimes
gives the same (within accuracy of about 50 MeV) predictions for lowest
 states.

 Indeed , variational calculations for Hamiltonian (2) with $\nu=\sigma
r$ gives for gluon subsystem energy (without Coulomb forces)
\be
E_{gl}=4.36\sqrt\sigma\approx1.7 GeV
\ee
String energy (6) for $L_{gl}=1$ is
\be
E_{gl}^{(0)}=\sqrt\sigma \sqrt{2\pi\sqrt2}=2.98\sqrt\sigma\approx1.2 GeV
\ee
is relatively small with respect to (8). Actually correction to regime
(6) is not small. Following the method described in [4] one can estimate
first correction to the string regime:
\be
\delta E_{gl}=\sqrt\pi \frac{5}{4}\frac{3^{1/5}}{2^{17/20}}\sqrt\sigma
\approx 1.53\sqrt\sigma
\ee
Adding (9) and (10) one obtains gluon subsystem energy calculated by
means of string regime formulaes:
\be
E_{gl}=4.51\sqrt\sigma\approx1.8 GeV
\ee
 As it is evident from (8) and (11) potential and string regimes match
smoothly and gives the same predictions for gluon energy of lowest
hybrids. This is a rather general property of the QCD string formalism
developed in [4,5] .
 It's instructive to estimate the deviations from regime (6) because of
quark correction. According to (1) interquark potential arise from the
terms:
\be
V(\rho)=\rho^2\left<\frac{d\sigma^2}{4}+\frac{\Pi L_{gl}^2}{12{(\mu_g+
2\Pi)}^2 r^4}-\frac{\Pi p_r^2}{6 \mu_g^2 r^2}\right >_{{\Psi}_{gl}}
\ee
after averaging over gluon wave function obtained from Hamiltonian (2).
It's easy to see that for regime (6) :
\be
d\sim 1/\sqrt L\ ;\ \frac {L^2}{\Pi r^4}\sim 1/\sqrt L\ ;\ \frac{\Pi}
{r^2} \sim 1/\sqrt L
\ee
For the last term in the straight-line ansatz [4] one can estimate :
\be
p_r^2/\mu_g^2 \sim 1/\root 4 \of L
\ee
Therefore the last term in (12) decreases most rapidly and there is no
effective interquark repulsion (cause there is no repulsion for $L_{gl}
=1$). Also it's evident that quark correction to regime (7) decreases
as $1/\root 4\of L$ (i.e. very slowly) and for regime (7) quark
correction to the gluon energy is about 100 MeV.

 Three Coulomb terms should be added to the full Hamiltonian:
\be
V_{Coul}=-\frac{3 \alpha_s}{2}\left(\frac{1}{r_1}+\frac{1}{r_2}\right)
+\frac 16 \frac{ \alpha_s }{\rho}
\ee
So,to take into account Coulomb forces one is to add the term $-3
\alpha_s /r$ to zeroth-approximation Hamiltonian (2). The explicit
form of quark  subsystem  Hamiltonian is:
\be
{\hat H}_{q\bar q}=\frac{{\vec p}^2}{m}+\Lambda \rho^2+\frac{1}{6}
\frac{\alpha_s}{\rho}
\ee
where oscillator term of the potential is derived in (12). We calculate
lowest eigenvalues of Hamiltonian (16) variationally and designate
them as $E_{q \bar q}$ .
 For the calculation of the absolute value of hybrid mass we use the
procedure proposed in [7,8] : the additive constant in the potential
is twice larger than for heavy-light system. Let us define the constant
term from masses of P-states of heavy-light system . The Hamiltonian
for heavy-light system coincides with hybrid Hamiltonian (2) after
replacing $2\sigma\to\sigma$. Let us neglect Coulomb for beginning.
Since energy $E_{gl}$ is proportional to $\sqrt\sigma$, heavy-light
gluon energy is $\sqrt 2$ smaller than for hybrid system . Therefore:
\be
M_{q\bar q g}=2M_{q\bar q}(L=1)+E_{gl}(1-\sqrt2)+E_{q\bar q}
\ee
Here $E_{gl}$ is gluon subsystem energy discussed above , $M_{q\bar q}$
is the experimental value of heavy-light state mass with $L=1$
 (2.42 GeV for D-meson ) . Actually heavy-light and hybrid
Hamiltonians differ by Coulomb terms and one is to use a function
$E_{gl}(\alpha_s)$ . Then
\be
M_{q\bar q g}=2M_{q\bar q}+E_{gl}(\alpha_s)-\sqrt2E_{gl}(4\alpha_s/9)
+E_{q\bar q}
\ee
{}From (18) it's clear that predicted hybrid mass in our model depends
on quark mass very slowly cause the only mass dependence is contained
only in the last term in (18) and this term is inversly proportional to
$\sqrt m$ . Numerical results for charmonium hybrids are listed in the
 Table I.
\begin{center}
{\bf Table I}

{\it Hybrid masses calculated by means of string and potential regimes
formulaes. The parameters are : $\sigma=0.18\ GeV^2 , \alpha_s=0.3$}
\end{center}
\begin{tabular}{llll}
& Quark mass,GeV&Potential,GeV&String,GeV\\
&$m_c=1.2$&4.209&4.134\\
&1.3&4.198&4.125\\
&1.5&4.178&4.117\\
&1.7&4.162&4.100\\
\end{tabular}

So, for charmonium hybrid mass we anticipate :
\be
M_{c \bar c g}=4.1\pm 0.1\ GeV
\ee
Since P-levels for B-mesons are still unknown we can't use (18) for
evaluation of bottomonium hybrid mass. Therefore we are to use the same
procedure as in [8] i.e. define additive constant from S-levels of
B-mesons. Hence, our predictions for $c \bar c g$ mass coincide with
predictions of [8] .

   We wish to discuss some aspects of flux-tube model [9] in comparison
with our results. First, there is no Regge-trajectories in flux-tube
model because it is purely nonrelativistic . Second , in the original
papers [9] small oscillations approximation for string excitations was
used. Numerical calculations in the frame of "one-bead flux-tube model"
[10] show that this is not the case. Moreover, rather general arguments
given in [8] prove that for (infinitely) heavy quarks gluon subsystem
oscillation is much larger than for quark subsystem. It's sufficient
that small string oscillations {\bf never} appear in the flux-tube
model ; hence , the approximation made in [9] is rather doubtful. Our
predictions for heavy hybrid mass coincide with flux-tube  model [9]
( and with lattice calculations [11]) , but it follows from the fact
that various prescriptions for additive constant were used .

  Quantum  numbers  of  our  hybrids  are  (see [7]):
\be
J^{PC}=0^{\mp +},1^{\mp +},2^{\mp +},1^{\mp -}
\ee
  Our hybrids can be produced directly in $e^+e^-$ collisions ( $J^{PC}
=1^{--}$ is allowed ), but appropriate width should be very small cause
hybrid itself could be produced through mixing between ordinary and
hybrid meson states . Of course, "gluish" channels like $p\bar p$
annihilation and highly excited quarkonium decays are preferrable.
Hybrid states should be very narrow  cause for $c\bar cg$ hybrids their
masses are close to open charm thresholds and decays are strongly
supressed by phase space ; for $b\bar b g$ hybrids decays to open bottom
are forbidden (strongly suppressed , at least).

     This research was made possible in part by Grant N J77100 from
International Science Foundation and Russian Government .The support
from Russian Fundamental Research Foundation, Grant N 93-02-14937 is
also acknowledged.

\end{document}